\documentclass[11pt]{article}
\usepackage{latexsym}
\usepackage{amsfonts,amssymb}
\usepackage{algorithm}
\usepackage[noend]{algorithmic}
\usepackage{amsmath,amssymb,amsthm,xspace,fullpage}
\usepackage{graphicx}
\usepackage{eucal}

\newtheorem{theorem}{Theorem}
\newtheorem{lemma}{Lemma}

\newtheorem{proposition}{Proposition}

\def\squarebox#1{\hbox to #1{\hfill\vbox to #1{\vfill}}}
\def\qed{\hspace*{\fill}        \vbox{\hrule\hbox{\vrule\squarebox{.667em}\vrule}\hrule}\smallskip}

 \newcommand{\bs}{\bigskip} 
  
 \newcommand{\hs}[1]{\hspace*{ #1 mm}}

\newcommand{\ignore}[1]{}

\newcommand{\prob}{\mathop\mathbf{Pr}}
\newcommand{\expe}{\mathop\mathbf{E}}

\newcommand{\ceil}[1]{\left\lceil #1 \right\rceil}
\newcommand{\floor}[1]{\left\lfloor #1 \right\rfloor}

\allowdisplaybreaks[1]

\begin{document}
%%%%%%%%%%%%%%%%%%
\pagestyle{plain}
\begin{center}
{\Large {\bf Improved Average Complexity for Comparison-Based Sorting}}
%\footnote{} 
\bs\\

{\sc Kazuo Iwama}$^1$ \hspace{5mm} 
{\sc Junichi Teruyama}$^2$$^3$ \hspace{5mm} 

\

{\small
$^1${RIMS, Kyoto University, Japan}; \\
{\tt iwama@kuis.kyoto-u.ac.jp} 

$^2${National Institute of Informatics, Japan}; \\
$^3${JST, ERATO, Kawarabayashi Large Graph Project, Japan}; \\
{\tt teruyama@nii.ac.jp} 

}
\end{center}
\bs

\begin{abstract}
This paper studies the average complexity on the number of
comparisons for sorting algorithms.  Its information-theoretic lower bound is
$n \lg n - 1.4427n + O(\log n)$. For many efficient algorithms, the
first $n\lg n$ term is easy to achieve and our focus is on the
(negative) constant factor of the linear term.  The current best value
is $-1.3999$ for the MergeInsertion sort.  Our new value is $-1.4106$,
narrowing the gap by some $25\%$.  An important building block of our
algorithm is ``two-element insertion,'' which inserts two numbers $A$
and $B$, $A<B$, into a sorted sequence $T$.  This insertion algorithm
is still sufficiently simple for rigorous mathematical analysis and 
works well for a certain range of the length of $T$ for which the simple
binary insertion does not, thus allowing us to take a complementary
approach with the binary insertion.
\end{abstract}

\section{Introduction}

A majority of existing sorting algorithms, including Bubble sort,
Quick sort, Heap sort, Merge sort and Insertion sort, are so-called
comparison-based sorts, in which our basic operation is a comparison of
two input numbers.  The complexity in terms of this measure, the
number of comparisons needed to obtain a sorted sequence, is an
obvious lower bound of the running time of the algorithm.  Thus it has
been a popular research topic in TCS to investigate its upper and lower
bounds for several sorting algorithms. 
Note that any sorting algorithm for $n$ elements can be described as a binary
decision tree having $n!$ leaves corresponding to all different
permutations of the $n$ elements.  The number of comparisons to obtain
one of them is the number of nodes on the path from the
root to the leaf corresponding to the sequence.  Therefore we have an
obvious lower bound, called an information-theoretic lower bound.
Namely, any sorting algorithm needs
$$\lceil \lg n! \rceil \approx n \lg n - 1.4427n + O(\log n).$$ 
comparisons in the worst case.

Usually
it is not very hard to obtain an upper bound of $n\lg n$. 
For instance, consider the BinaryInsertion sort
that increases the length of the sorted sequence one by one using
binary insertion.  Obviously we have $n-1$ steps and each of them consists of 
at most $\ceil{\lg n}$ comparisons 
(and much less for most of the steps).
Thus our interest naturally comes to the constant factor for the
linear term in $n$.  Unfortunately, however, its analysis does not
seem so easy and our knowledge is quite limited.
For instance, it is at most $-0.91$ for Merge (and similar other) 
sort~\cite{Knuth98,Steinhaus50} and
the current best one is $-1.32$ for MergeInsertion sort obtained by
Ford and Johnson more than five decades ago~\cite{FJ59}.  

Our interest in this paper is the {\em average-case complexity}
on the number of comparisons, which should be easier to obtain than
the worst-case complexity.  In fact we do have a number of better
results; $-1.26$ for Merge sort~\cite{Knuth98}, $-1.38$ 
for BinaryInsertion sort, and most
recently $-1.3999$ for MergeInsertion sort~\cite{EW14}. 
Notice that 1.3999 is some
96.98\% of 1.4427, but there still exists a gap and seeking the exact
bound for this fundamental problem should be an important research goal. 

{\bf Our Contribution}~~ We achieve 1.4034 by a new algorithm {\sc (1,2)Insertion}.
Furthermore it is shown that the constant is improved to $-1.4106$ by
combining the new algorithm with the MergeInsertion sort.  Thus we have
narrowed the previous gap between 1.3999 and 1.4427 by some 25\%.  Our
new algorithm is based on binary insertion.
Recall that the BinaryInsertion sort repeats a binary insertion of a new
item into a sorted sequence of length $i-1$ for $i=2$ to $n$.  Here
the performance of binary insertion itself is optimal because it
constitutes an optimal decision tree of height $\lceil \lg i \rceil $. 
However, if $i$ is not a power of two, this tree is not completely 
balanced, i.e.,
there is a difference of one comparison due to the position of the
inserted element.  This small difference in each step accumulates during the 
repeated steps
and finally creates a large imbalance.
This is the reason for its relatively poor performance.

Our idea is to use a binary insertion if $i$ is close to a power of two 
and to use what we call a ``two-element merge,'' or {\sc 2Merge} otherwise.
{\sc 2Merge} merges a two-element sequence $(A, B)$, $A<B$, with
a sorted sequence $T$ of length $i-2$ to obtain a sorted sequence
of length $i$.  We first insert $A$ using a kind of binary search,
meaning $A$ is compared with an element in $T$ whose position is
approximately $1/3$ from the smallest.  If $A$ falls into the first
third of $T$, then we use a standard (with a bit of care) 
binary search, called {\em right-heavy binary search} or {\em RHBS}.
The key thing here
is that the original ``bad'' $i$ changes to a ``good'' $i'$ in this binary insertion. 
If $A$ falls into the right part of $T$, we simply recurse.  
Then, we insert $B$ into $T$ by using a standard binary search.
Thus we can reduce the imbalance of each step of insertion, which contributes
to the better bound for the whole sorting.  

Due to~\cite{EW14}, the performance of MergeInsertion differs a lot for
different $n$ and it hits a best peak when $n$ is about one third from
the previous power-of-two number, which achieves around $-1.415$. This is much
better than our {\sc (1,2)Insertion} (but, unfortunately, it quickly gets
worse as $n$ leaves the best position and ends up with $-1.3999$ for a
roughly power-of-two $n$).  Thus here is a natural idea: For a given
sequence $X$ of length $n$ that is bad for MergeInsertion, select the
largest value $n'$ that is less than $n$ and is good for
MergeInsertion.  Then we use MergeInsertion sort for a length $n'$
subsequence of $X$ and insert the remaining elements using 
{\sc (1,2)Insertion}, which in fact gives us $-1.4106$.

{\bf Related Work}~~ The idea of inserting two elements into a sorted sequence is not new. 
\cite{HL71} and \cite{TAB86} claimed two exactly optimal algorithms for such 
a merge operation in terms of the worst-case complexity~\cite{HL71} and in terms of the
average-case complexity~\cite{TAB86}.
Unfortunately, both algorithms are a bit
involved and their performance analysis did not give closed formulas
for the complexity.  Our {\sc 2Merge} is probably not exactly optimal,
but is sufficiently simple for rigorous mathematical analysis.

The analysis of the BinaryInsertion sort by Edelkamp
and Wei\ss~\cite{EW14} gives many hints to our new analysis. They show that
the average number of comparisons is
\begin{equation}\label{eq:ins1}
\ceil{\lg i} + {\cal B}(i), \text{ where } {\cal B}(i) = 1 - \frac{2^{\ceil{\lg i}}}{i}
\end{equation}
for a single insertion and is
\begin{eqnarray}
\sum_{i=1}^{n}\left(\ceil{\lg i} + {\cal B}(i)\right)
&=& n \lg n + \left( 1 - \lg p_n - \frac{1 + \ln (4p_n)}{p_n} \right)\\
&<& n \lg n - 1.386n \nonumber
\end{eqnarray}
for the entire BinaryInsertion sort, where $p_n=\frac{n}{2^{\ceil{\lg n}}}$ is a parameter indicating the deviation from a power of two.
Edelkamp and Wei\ss~\cite{EW14} also includes a nice survey on this topic.

Although we have few results on the worst-case complexities for
asymptotically large $n$, we do have a rather rich literature for small $n$'s.
For instance, the information-theoretic bound (actually its ceiling) cannot be achieved
by any comparison-based sorting for $12\leq n \leq 15$.  The
MergeInsertion sort achieves a matching upper bound for $1\leq n \leq
15$, but $n=16$ is still open, namely there is a gap of one between
the lower and upper bounds (45 and 46, resp.) for the exact number of
necessary comparisons. It is also known that MergeInsertion is not optimal for
some $n$'s, for instance, for $n=47$. 
See~\cite{ayala2007,FJ59,mana79,mana89,mon81,pec2002,pec2004,pec2007,well66}
for these results. 

{\bf Notations and Assumptions.}
Our sorting algorithm takes a {\em sequence} of all different $n$ {\em
elements} as input.  An {\em average complexity} (or simply {\em
complexity}) of a sorting algorithm {\sc Alg} is the expected number
of comparisons {\sc Alg} executes to sort each of $n!$ different sequences
of length $n$.  Note that the complexity of all sorting algorithms in
this paper is written as $n \lg n + cn + O(\log n)$ for some negative constant $c$. 
It is important to mention that the value of $c$, that is our main issue, 
periodically changes depending on
$n$ usually and we are interested in its worst (largest) value for
asymptotically large $n$, unless otherwise stated.  We exploit the
$O(\log n)$ term to make analysis simpler.  In particular we assume,
without loss of generality, that $n$ is always even throughout this
paper.  Also, when summing up a cost function
$f(i)$ for $i=$ 1 to $n$, an $O(1/i)$ term in $f$ is not important.
For notation, we
write $x=y \pm z$ if $|x-y|<z$, where $z$ may be a big-O
notation like $x=y \pm O(z)$.
We may denote a sequence of one element $(s_1)$ 
by simply $s_1$.

\section{Our Algorithm and Its Analysis}\label{sec:ins2}

\begin{algorithm}[b]
\caption{ {\sc(1,2)Insertion}($S$)}
{\bf Input:} A (unsorted) sequence $S=(s_1, s_2, ..., s_n)$, where $n$ is even.\\
{\bf Output:} Sorted sequence\\
Step~1: If $n=2$, then sort $(s_1, s_2)$ with a single comparison.\\
Step~2: Sort $S'=(s_1, ..., s_{n-2})$ by {\sc (1,2)Insertion} to obtain $T'$.\\
Step~3: If $p_n \in [0.5511, 0.888]$ then insert $s_{n-1}$ and $s_{n}$ into $T'$ by calling {\sc 2Merge}$(s_{n-1}, s_{n}, T')$.
Otherwise insert $s_{n-1}$ into $T'$ by {\sc RHBS} and then $s_{n}$ by {\sc RHBS}.
\end{algorithm}
\begin{algorithm}[!b]
\caption{{\sc 2Merge}($A$, $B$, $T$)}
\begin{algorithmic}
\REQUIRE $A$ and $B$ are numbers and $T=(t_1, t_2, ..., t_{i-2})$ is a sorted sequence
such that $i$ is even and $i\geq 4$.
\ENSURE Sorted sequence of length $i$.
\end{algorithmic}
Step 1. Compare $A$ and $B$ and swap them if $A>B$.\\
Step 2. Let $\alpha(r) = 1 - 2^{-r/2}$. For $r=1,2,\ldots,$ up to $2\lg i$,
compare $A$ with $t_{\ceil{\alpha(r)i}}$ and go to Step 3 if $A<t_{\ceil{\alpha(r)i}}$.\\
Step 3. Insert $A$ to $(t_{\ceil{\alpha(r-1)i}}+1, \ldots, t_{\ceil{\alpha(r)i}}-1)$ using {\sc RHBS}.
Suppose that $A$ falls between $t_{\ell}$ and $t_{\ell+1}$.\\
Step 4. Insert $B$ to $(t_{\ell+1}, \ldots, t_{i-2})$ using {\sc RHBS}.
\end{algorithm}
\begin{algorithm}[!b]
\caption{{\sc RHBS}($A$, $T$)}
{\bf Input:}  $A$ is a number and $T=(t_1, t_2, ..., t_i)$ is a sorted sequence.\\
{\bf Output:} Sorted sequence of length $i+1$.\\
Step 1. If $i\leq 3\times 2^{\ceil{\lg (i+1)}-2}-1$, then let set $d:=2^{\ceil{\lg (i+1)}-2}$.
Otherwise, let set $d:=i-2^{\ceil{\lg (i+1)}-1}+1$.\\
Step 2. Let $T_1=(t_1, \ldots, t_{d-1})$ and $T_2=(t_{d+1}, \ldots, t_{i})$.\\
Step 3. Compare $A$ with $t_d$.
If $A < t_d$, return {\sc RHBS}$(A, T_1) \circ t_d \circ T_2$.
Otherwise, return $T_1 \circ t_d \circ${\sc RHBS}$(A, T_2)$.
\end{algorithm}

See Algorithms~1, 2 and~3.
The main algorithm is Algorithm~1.
Note that Algorithm~2 is improved in the next section and Algorithm~1 is
combined with MergeInsertion in Section~\ref{sec:comb}.
For a given sequence $S=(s_1, s_2, \ldots, s_{n-1}, s_n)$ with an even $n$, 
{\sc (1,2)Insertion} works in Round 0, Round 2, ... up to Round $n-2$.
In Round 0, $s_1$ and $s_2$ are sorted by a single comparison to make a
sorted sequence $T_0$ of length two.  In Round $i$, $s_{i+1}$ and
$s_{i+2}$ are inserted into $T_{i-2}$ obtained in Round $i-2$ by using 
(i) a single call of {\sc 2Merge} or (ii) two calls of {\sc RHBS},
depending on the value $i$. Recall that we wish to obtain the average
complexity for all different $n!$ sequences, in other words, we wish to obtain
the expected number of comparisons assuming that each $S$ appears
uniformly at random.  It then turns out that  
we can also assume that the position of 
$s_{i+1}$ (and that of $s_{i+2}$ also) in each round is uniformly at
random in the different $i+1$ positions of $T_{i-2}$ that includes $i$
elements. Thus the overall average complexity is a simple sum of the
average complexity of each round.

We first make an analysis of {\sc 2Merge}. 
Note that {\sc 2Merge} uses {\sc RHBS} which stands for Right-Heavy Binary Search.
Note that the number, say $q$, of comparisons to insert $A$ into a sequence
$T=(t_1, \ldots t_i)$ is $q_0=\ceil{\lg(i+1)}-1$ or $q_0+1$ if
we use the standard binary search.
The feature of RHBS is that if $q=q_0+1$ for
some $A$, then $q=q_0+1$ for any $A'$ such that $A'>A$, in
other words, the number of comparisons is monotone.  This is
easily realized by selecting $t_d$ (to be compared with $A$) in each recursion phase such
that either the number of $T$'s elements that is smaller than $t_d$ 
or the number of those that is larger than $t_d$ 
be (a power of two)$-1$. Suppose for instance $8\leq i \leq 15$. 
Then if $i$ is 11 or less, then the first comparison is with $t_4$ and if $i$ is
12 or more, then the first comparison is with $t_{i-7}$.
There would be no merit of this structure if the position of $A$
is uniformly distributed.  However, if small $A$'s are more likely
than large $A$'s, there is an obvious advantage and that provides a
real merit in {\sc 2Merge}.
Notice that even if our improvement in each step is a small constant, 
something like 0.1,
that constant significantly affect the value of our constant factor of the linear term.

In Step 2, we determine the range of the smaller element $A$.  If the
condition there ($A < t_{\ceil{\alpha(r)i}}$) is met for $r=1$, then
the range is $(t_1, \ldots, t_{\ceil{(1-1/\sqrt{2})i}-1})$, where
$(1-1/\sqrt{2}) \approx 0.2929$.  In general, the range is
$(t_{\ceil{\alpha(r-1)i}+1}, \ldots,t_{\ceil{\alpha(r)i}-1})$ for an integer $r \geq 1$, 
and we wish to compute the average complexity of Step 3, i.e., the average number
of comparisons to insert $A$ into this range.  
Here we have two technical issues: (i) We introduce a parameter $w_r$ and let
$w_r := (\sqrt{2}-1)2^{-r/2}i$.  Note that $w_r$ is somehow related to
the size of the above range but it may not be integral.  The idea is
that the complexity does not differ significantly if the size of the range
differs by a small constant and approximating the size by $w_r$ makes
our job much easier.  (ii) Although the positions of $A$ and $B$ are
uniformly at random, we now know that $A<B$.  Therefore the
probability that $A$ falls between $t_{\ell-1}$ and $t_{\ell}$ under
the condition that $A<B$ is $(i-\ell)/{\binom{i}{2}}$.  We also extend
the definition of $p_x=\frac{x}{2^{\ceil{\lg x}}}$ for a noninteger $x$.

\begin{lemma}\label{lem:step3}
Suppose that $A$ is to be inserted to $(t_{\ceil{\alpha(r-1)i}+1},
\ldots,t_{\ceil{\alpha(r)i}-1})$ for an even $i$.
Then {\sc 2Merge} requires 
\begin{equation}\label{eq:fix_r}
{\cal A}(r)= \ceil{\lg w_r} + 7-4\sqrt{2} - \frac{10-6\sqrt{2}}{p_r} + 
\frac{3-2\sqrt{2}}{p_r^2}\pm O\left(\frac{2^{r/2}}{i}\right)
\end{equation}
comparisons on average at Step~3.  Furthermore, the expected value of ${\cal A}(r)$ is 
\begin{eqnarray*}
\prob[r=1]{\cal A}(1)+\prob[r=2]{\cal A}(2)+\cdots = 
\ceil{\lg i} + {\cal T}(i), 
\end{eqnarray*}
where 
\begin{eqnarray}\label{eq:T_i}
{\cal T}(i) = 5-4\sqrt{2} - \frac{1}{p_i} + \frac{1}{6 p^2_i} + 
\left\{
\begin{array}{ll}
-\frac{1}{6 p_i}  - \frac{1}{16p^2_i} -\frac{2}{3} \:\:& p_i \in (1/2, \frac{1+\sqrt{2}}{4}], \\
-\frac{\sqrt{2}}{3 p_i} -\frac{1}{3} & p_i \in (\frac{1+\sqrt{2}}{4}, \frac{2+\sqrt{2}}{4}], \\
-\frac{4}{3p_i} + \frac{1}{4p^2_i} + \frac{1}{3} & p_i \in (\frac{2+\sqrt{2}}{4}, 1].
\end{array}
\right.
\end{eqnarray}
\end{lemma}

See Section~\ref{sec:lemma1} for the proof.
Now we are going to Step 4 to insert $B$ 
and here is our analysis (see Section~\ref{sec:lemma2} for its proof).

\begin{lemma}\label{lem:step4}
For an even $i$, 
{\sc 2Merge} requires
\[
\ceil{\lg(i-1)} + 1 - \frac{2}{p_i} + \frac{1}{3p_i^2} + O(1/i)
\]
comparisons on average at Step~4.
\end{lemma}

The entire complexity of {\sc 2Merge} is  the sum of these two
quantities in Lemmas 1 and 2 and another two values; $(+1)$ for
comparing $A$ and $B$ at Step~1 and the one for the expected number of
comparisons in Step 2 that is $2 \pm O(1/i)$ (see Appendix~\ref{app:expe_r}).
Thus the complexity of {\sc 2Merge} is 
\[
\ceil{\lg{i}} + \ceil{\lg(i-1)}+ {\cal U}(i) +O(1/i)
\]
where (${\cal T}(i)$ is equation~{(\ref{eq:T_i})})
\begin{equation}
{\cal U}(i) = 1 + {\cal T}(i) - \frac{2}{p_{i-1}} + \frac{1}{3p_{i-1}^2}.
\end{equation}
Since this is the complexity for inserting two elements, the
complexity for a single insertion can be regarded as a half of it, or
\begin{equation}\label{eq:ins2_per1}
\ceil{\lg{i}} + {\cal U}(i)/2 +O(1/i).
\end{equation}
It then turns out that by comparing this value with 
(\ref{eq:ins1}) of the BinaryInsertion, 
{\sc 2Merge} is better than BinaryInsertion for $0.5511 < p_i < 0.888$.
(Note that this range is obtained by a numerical calculation.)
Thus we use {\sc 2Merge} for this range of $p_i$
and {\sc RHBS} for the other range.
In summary our one step complexity is 
\begin{eqnarray*}
\ceil{\lg{i}} + 
\left\{
\begin{array}{ll}
{\cal B}(i) & p_i \in (1/2, 0.5511] \\
{\cal U}(i)/2 +O(1/i) \quad & p_i \in (0.5511, 0.888] \\
{\cal B}(i) &p_i \in (0.888, 1]
\end{array}
\right.
\end{eqnarray*}

By simple calculation, this is rewritten by 
\begin{eqnarray}
\ceil{\lg i} + {\cal D}(p_i)
\end{eqnarray}
where 
\begin{eqnarray}
{\cal D}(p_i)=
\left\{
\begin{array}{ll}
1 - \frac{1}{p_i} & p_i \in \left(1/2, 0.5511\right], \\
\frac{25}{6}-2\sqrt{2} - \frac{19}{12p_i} + \frac{7}{32p_i^2} \quad &
p_i \in \left(0.5511, \frac{1+\sqrt{2}}{4}\right], \\
\frac{13}{3} -2\sqrt{2} - \frac{9+\sqrt{2}}{6 p_i} + \frac{1}{4p_i^2} & 
p_i \in \left(\frac{1+\sqrt{2}}{4}, \frac{2+\sqrt{2}}{4}\right], \\
\frac{14}{3}-2\sqrt{2} - \frac{13}{6p_i} + \frac{3}{8p_i^2} &
p_i \in \left(\frac{2+\sqrt{2}}{4}, 0.888\right],\\
1 - \frac{1}{p_i} & p_i \in \left(0.888, 1\right].
\end{array}
\right.
\end{eqnarray}
Now by using the trapezoidal rule, we have
\begin{eqnarray*}
\sum_{i=1}^{n}{\cal D}(p_i)=
2^{\ceil{\lg n}} \times \left\{
\int_{1/2}^{1}{\cal D}(x)dx
 + \int_{1/2}^{p_n}{\cal D}(x)dx
 \right\} + O(\log n)
\end{eqnarray*}
and the following theorem. (See Appendix~\ref{app:sum_int} for details.)

\begin{theorem}
The complexity of {\sc (1,2)Insertion} is at most $n \lg n - 1.40118n$.
\end{theorem}

%%%%%%%%%%%%%%%%%%%%%%%%%
% Sub-Section : Lemma1
%%%%%%%%%%%%%%%%%%%%%%%%%
\subsection{Proof of Lemma~\ref{lem:step3}}\label{sec:lemma1}

We first prove formula~{(\ref{eq:fix_r}).
By the assumption of the lemma, we call {RHBS}$(A, (t_{\ell_1+1}, \ldots,
t_{\ell_2-1}))$, where 
$\ell_1 = \ceil{(1-2^{-(r-1)/2})i}$ and $\ell_2 =
\ceil{(1-2^{-r/2})i}$.
For an integer $\ell$, let $E_\ell$ denote the event that $A$ falls
between $t_{\ell-1}$ and $t_{\ell}$. Also $F$ denotes the event 
that $A$ is inserted between $t_{\ell_1}$ and $t_{\ell_2}$, namely
$F = \bigcup_{\ell=\ell_1+1}^{\ell_2}E_\ell.$
Let $w = \ell_2 - \ell_1$ and $z = 2i-\ell_1 - \ell_2 - 1$.
Since $\prob[E_\ell] = \frac{i-\ell}{\binom{i}{2}}$, we have 
\begin{eqnarray*}
\prob[F] 
= \sum_{\ell=\ell_1+1}^{\ell_2}\frac{i-\ell}{\binom{i}{2}} 
= \frac{w \cdot z}{2\binom{i}{2}}
\text{ and }
\prob[E_\ell \mid F] = \frac{\prob[E_\ell]}{\prob[F]} = \frac{2(i-\ell)}{w \cdot z}.
\end{eqnarray*}

Let $k = 2^{\ceil{\lg w}} - w$. By its monotonicity,
RHBS requires $\ceil{\lg w} - 1$ comparisons if $t_{\ell_1} < A < t_{\ell_1 + k}$,
and requires $\ceil{\lg w}$ comparisons otherwise.
Therefore, the average number of comparisons is
$\ceil{\lg w} - \sum_{\ell=\ell_1 + 1}^{\ell_1 + k}\prob[E_\ell \mid F]$,
we need to calculate the summation
$\sum_{\ell=\ell_1 + 1}^{\ell_1 + k}\prob[E_\ell \mid F] = 
\sum_{\ell=\ell_1 + 1}^{\ell_1 + k}\frac{2(i-\ell)}{w\cdot z}$.
Observing that $k/w = 1/p_w -1$, we have
\begin{eqnarray}
\sum_{\ell=\ell_1 + 1}^{\ell_1 + k}\frac{2(i-\ell)}{w\cdot z}
&=& 
\frac{k}{w} \cdot \frac{2i-2\ell_1 - k-1}{z}
\nonumber \\
&=& 
\frac{k}{w} \cdot \frac{z + w - k}{z} \quad\quad(\because 2i - 2\ell_1 - 1 = w + z)
\nonumber \\
&=& 
\frac{k}{w} \cdot \left(1 + \frac{w - k}{w} \cdot \frac{w}{z}\right) 
\nonumber \\
&=& 
\frac{1}{p_w} - 1 + \left( - 2 + \frac{3}{p_w} - \frac{1}{p^2_w} \right)  \cdot \frac{w}{z} \label{eq:sum:lemma1}
\end{eqnarray}

Since $\ell_1 = \ceil{(1-2^{-(r-1)/2})i}$ and $\ell_2 =
\ceil{(1-2^{-r/2})i}$.
we have $w = 2^{-r/2}(\sqrt{2}-1)i \pm 1$ and $z=2^{-r/2}(\sqrt{2}+1)i \pm1$.
Observe the value $\frac{w}{z}$ is close to $3-2\sqrt{2}$, in fact
the difference is bounded as
\begin{eqnarray*}
\left| 3-2\sqrt{2} - \frac{w}{z} \right|
&<&
\frac{4-2\sqrt{2}}{z}
<
\frac{2^{r/2}}{i} \quad\left(\because r \leq 2\lg i\right).
\end{eqnarray*}
Therefore, because $k/w = 1/p_w - 1$ and $ - 2 + \frac{3}{p_w} -
\frac{1}{p^2_w} \leq \frac{1}{4}$, (\ref{eq:sum:lemma1}) continues as
\begin{eqnarray*}
(*) &=&\frac{1}{p_w} - 1 + \left( - 2 + \frac{3}{p_w} - \frac{1}{p^2_w} \right)  \cdot \left( 3-2\sqrt{2} \pm \frac{2^{r/2}}
{i} \right) \\
&=& -7+4\sqrt{2} + \frac{10-6\sqrt{2}}{p_r} - \frac{3-2\sqrt{2}}{p_r^2} \pm \frac{2^{r/2}}{4i}
\end{eqnarray*}
Thus, the average number of comparisons is
\begin{equation}\label{eq:ave_w}
\ceil{\lg w} + 7-4\sqrt{2} - \frac{10-6\sqrt{2}}{p_w} + \frac{3-2\sqrt{2}}{p_w^2} \pm \frac{2^{r/2}}{4i}.
\end{equation}

As mentioned before the statement of the lemma, we wish to replace $\lg w$ by $\lg w_r$, 
since there is no obvious way of treating the ceiling of the former
that includes another ceilings for $w$.
Now, recall that $w_r = 2^{-r/2}(\sqrt{2}-1)i$ and $p_r = \frac{w_r}{2^{\ceil{\lg w_r}}}$.
We show that it is possible to simply replace $\ceil{\lg w}$ by
$\ceil{\lg w_r}$ almost as it is:
(i) If $\ceil{\lg w} = \ceil{\lg w_r}$ holds, then 
because $|\frac{1}{p_w} - \frac{1}{p_r}| = O\left(\frac{2^{r/2}}{i}\right)$, 
it is enough to replace the last (error)
term with $O\left(\frac{2^{r/2}}{i}\right)$.
(ii) Otherwise suppose that $\ceil{\lg w} \neq \ceil{\lg w_r}$.
Since $|w-w_r|<1$ and $w$ is an integer, 
$w$ must be a power of two and $\ceil{\lg w_r}$ must be $\ceil{\lg w}+1$.
It then follows that $p_w = 1$ and we can write that $1/p_r = 2 - \epsilon$, 
where $|\epsilon| = \frac{2|w_r-w|}{w_r} < (2\sqrt{2}+2) \cdot
2^{r/2}/i$.
Substituting $p_w=1$, (\ref{eq:ave_w}) becomes
\[
\ceil{\lg w} \pm \frac{2^{r/2}}{4i}.
\]
Substituting $\ceil{\lg w_r} = \ceil{\lg w} + 1$ and $1/p_r = 2-\epsilon$, (\ref{eq:fix_r}) becomes
\begin{eqnarray*}
\ceil{\lg w} + (2\sqrt{2}-2)\epsilon +(3-2\sqrt{2})\epsilon^2 \pm O\left(\frac{2^{r/2}}{i}\right)
=
\ceil{\lg w} \pm O\left(\frac{2^{r/2}}{i}\right).
\end{eqnarray*}
Therefore (\ref{eq:ave_w}) i.e., the value we want to obtain
can be replaced by (\ref{eq:fix_r}) with the error term.
Thus the former part of lemma is proved.

For formula (4) we need to give the average values of 
$\ceil{\lg w_r}$, $1/p_r$ and $1/p_r^2$.  Since $\lg i = \ceil{\lg i} + \lg
p_i$ and $\ceil{x} = - \floor{-x}$ for any value $x$, we have
\begin{eqnarray*}
\ceil{\lg w_r} 
= \ceil{\lg i + \lg(\sqrt{2}-1) - r/2}
= \ceil{\lg i} - \floor{r/2 - \lg(p_i(\sqrt{2}-1))}.
\end{eqnarray*}
Also, for any value $x$ and integer $m$, we have
\[
\floor{r/2 + x} =
\left\{
\begin{array}{ll}
\floor{r/2} + m \:\:& x \in [m,m+1/2), \\
\ceil{r/2} + m \:\:& x \in [m+1/2,m+1). \\
\end{array}\right.
\]
Let 
$$c_r(p_i) = \floor{r/2 - \lg (p_i(\sqrt{2}-1))}.$$
Then since $\lg (p_i(\sqrt{2}-1)) \in (-2.5, -1)$, we have
\[
c_r(p_i) :=
\left\{
\begin{array}{ll}
\floor{r/2} + 2 \:\:&p_i \in (1/2, \frac{1+\sqrt{2}}{4}], \\
\ceil{r/2} + 1 \:\:&p_i \in (\frac{1+\sqrt{2}}{4}, \frac{2+\sqrt{2}}{4}], \\
\floor{r/2} + 1 \:\:&p_i \in (\frac{2+\sqrt{2}}{4}, 1].
\end{array}
\right.
\]
We have the following lemmas about the expected values of $\lceil r/2 \rceil$ and $\lfloor r/2 \rceil$.
(See Appendix~\ref{app:exp_ceil} for the proof.)
\begin{lemma}\label{lem:expect_ceilfloor}
$\expe[\floor{r/2}] = 2/3 \pm O(1/i)$ and $\expe[\lceil r/2 \rceil] = 4/3 \pm O(1/i)$.
\end{lemma}
This lemma implies
\[
\expe[\ceil{\lg w_r}] 
= \ceil{\lg i} \pm O(1/i) - \left\{
\begin{array}{ll}
8/3 \quad& p_i \in (1/2, \frac{1+\sqrt{2}}{4}], \\
7/3 & p_i \in (\frac{1+\sqrt{2}}{4}, \frac{2+\sqrt{2}}{4}], \\
5/3 & p_i \in (\frac{2+\sqrt{2}}{4}, 1].
\end{array}
\right.
\]

Similarly, we can obtain the expected value of $1/p_r$ and $1/p^2_r$ as follows.
(See Appendix~\ref{app:exp_pr} for the proof.)
\begin{lemma}\label{lem:pw}
\[
\expe[1/p_r] 
=
\left\{
\begin{array}{ll}
\frac{3\sqrt{2}+5}{12p_i} & p_i \in (1/2, \frac{1+\sqrt{2}}{4}] \\
\frac{3+2\sqrt{2}}{6p_i} &  p_i \in (\frac{1+\sqrt{2}}{4}, \frac{2+\sqrt{2}}{4}] \\
\frac{3\sqrt{2}+5}{6p_i} & p_i \in (\frac{2+\sqrt{2}}{4}, 1]
\end{array},
\right.
\:\:
\expe\left[1/p_r^2\right] = \left\{
\begin{array}{ll}
\frac{5(3+2\sqrt{2})}{48p^2_i} & p_i \in (1/2, \frac{1+\sqrt{2}}{4}] \\
\frac{3+2\sqrt{2}}{6p^2_i} &  p_i \in (\frac{1+\sqrt{2}}{4}, \frac{2+\sqrt{2}}{4}] \\
\frac{5(3+2\sqrt{2})}{12p^2_i} & p_i \in (\frac{2+\sqrt{2}}{4}, 1]
\end{array}\right..
\]
\end{lemma}

Adding all those values, we can obtain (4) and the lemma is proved.

%%%%%%%%%%%%%%%%%%%%%%%%%
% Sub-Section : Lemma2
%%%%%%%%%%%%%%%%%%%%%%%%%
\subsection{Proof of Lemma~\ref{lem:step4}}\label{sec:lemma2}

If $t_{\ell-1}<A<t_{\ell}$,
{\sc RHBS}($B$, ($t_{\ell}, \ldots, t_{i-2}$)) is called.
Since there are $i-\ell$ positions for the insertion, 
the average number of comparisons
is $\lceil \lg(i-\ell) \rceil + 1 - \frac{2^{\lceil \lg(i-\ell) \rceil}}{i-\ell}$
using the formula for the standard binary search
($B$ 's position is uniformly distributed).
Because $\Pr[t_{\ell-1}<A<t_{\ell}] = \frac{i-\ell}{\binom{i}{2}}$,
its expected value is 
\begin{eqnarray*}
\sum_{\ell=1}^{i-1} 
\left\{\lceil \lg(i-\ell) \rceil + 1 - \frac{2^{\lceil \lg(i-\ell) \rceil}}{i-\ell}\right\} \cdot 
\frac{i-\ell}{\binom{i}{2}} 
=1 + \left( \sum_{t=1}^{i-1} t\ceil{\lg t} - \sum_{t=1}^{i-1} 2^{\ceil{\lg t}} \right)\times \frac{1}{\binom{i}{2}}.
\end{eqnarray*}

Let $m = \ceil{\lg(i-1)}$. Then the first sum is
\begin{eqnarray*}
\sum_{t=1}^{i-1} t\lceil \lg t \rceil &=&
\sum_{d=1}^{m-1} \sum_{k=2^{d-1} + 1}^{2^{d}} kd + \sum_{t=2^{m-1}+1}^{i-1}tm \\ 
&=&
\sum_{d=1}^{m-1}\left\{ \frac{3}{8}d4^d + \frac{1}{4} d 2^d\right\} + \frac{(i+2^{m-1})(i-1-2^{m-1})m}{2} \\ 
&=&
\frac{3}{8} \cdot \left( \frac{1}{3}m4^m - \frac{4}{9}4^{m} + \frac{4}{9} \right)
+ \frac{1}{4} \cdot \left( m2^{m} - 2^{m+1} + 2 \right)
+ m \cdot \left( \binom{i}{2} - 2^{m-2} - 2^{2m-2} \right) \\ 
&=&
m \binom{i}{2} - \frac{1}{6}4^m - \frac{1}{2}2^{m} + \frac{2}{3}
\end{eqnarray*}
where we used $\sum_{d=1}^{m-1}d\cdot k^d = \frac{m k^m}{k-1} - \frac{k^{m+1}}{(k-1)^2} + \frac{k}{(k-1)^2}$
for the third equation.
Next, we calculate the second sum:
\begin{eqnarray*}
\sum_{t=1}^{i-1}2^{\ceil{\lg t}} &=& 
1 + \sum_{d=1}^{m-1} 2^{2d-1} + \sum_{t=2^{m-1}+1}^{i-1} 2^{m} \\ 
&=&
1 + \frac{4^{m}}{6} - \frac{2}{3} + (i - 2^{m-1}-1) \cdot 2^{m} \\
&=&
i \cdot 2^m - \frac{1}{3}4^{m}  - 2^m + \frac{1}{3}
\end{eqnarray*}

Therefore, 
\begin{eqnarray*}
\left( \sum_{t=1}^{i-1} t\ceil{\lg t} - \sum_{t=1}^{i-1} 2^{\ceil{\lg t}} \right)\times \frac{1}{\binom{i}{2}}
&=&
\frac{m \binom{i}{2} - i \cdot 2^m + \frac{1}{6}4^m + \frac{1}{2}2^{m} + \frac{1}{3}}{\binom{i}{2}} \\
&=& \ceil{\lg(i-1)} - \frac{2}{p_{i-1}} + \frac{1}{3p_{i-1}^2} + O(1/i).
\end{eqnarray*}
Since $i$ is even, $\frac{1}{p_{i-1}} = \frac{1}{p_{i}} + O(1/i)$,
which completes the proof.
\qed

%%%%%%%%%%%%%%%%%%%%%%%%%%%%%%%%%%%%%%%%%%
% Section : Improved Algorithm
%%%%%%%%%%%%%%%%%%%%%%%%%%%%%%%%%%%%%%%%%%
\section{Improvement of {\sc 2Merge}}

\begin{algorithm}[t]
\caption{ {\sc(1,2)Insertion*}($S$)}
{\bf Input:} A (unsorted) sequence $S=(s_1, s_2, ..., s_n)$, where $n$ is even.\\
{\bf Output:} Sorted sequence\\
Step~1: If $n=2$, then sort $(s_1, s_2)$ with a single comparison.\\
Step~2: Sort $S'=(s_1, ..., s_{n-2})$ by {\sc (1,2)Insertion*} to obtain $T'$.\\
Step~3: If $p_n \in \left[ \frac{3}{4} - \frac{\sqrt{6}}{12}, \frac{3}{4} + \frac{\sqrt{3}}{12}\right]$
 then insert $s_{n-1}$ and $s_{n}$ into $T'$ by calling {\sc 2Merge}$(s_{n-1}, s_{n}, T')$.
Otherwise insert $s_{n-1}$ into $T'$ by {\sc RHBS} and then $s_{n}$ by {\sc RHBS}.
\end{algorithm}
\begin{algorithm}[t]
\caption{{\sc 2Merge*}($A$, $B$, $T$)}
\begin{algorithmic}
\REQUIRE $A$ and $B$ are numbers and $T=(t_1, t_2, ..., t_{i-2})$ is a sorted sequence
such that $i$ is even and $i\geq 4$.
\ENSURE Sorted sequence of length $i$.
\end{algorithmic}
Step 1. Compare $A$ and $B$ and swap them if $A>B$.\\
Step 2. Define $\alpha(r, p_i)$ as Equation~{(\ref{eq:alpha*})}. For $r=1,2,\ldots,$ up to $2\lg i$,
compare $A$ with $t_{\ceil{\alpha(r)i}}$ and go to Step 3 if $A<t_{\ceil{\alpha(r)i}}$.\\
Step 3. Insert $A$ to $(t_{\ceil{\alpha(r-1)i}}+1, \ldots, t_{\ceil{\alpha(r)i}}-1)$ using {\sc RHBS}.
Suppose that $A$ falls between $t_{\ell}$ and $t_{\ell+1}$.\\
Step 4. Insert $B$ to $(t_{\ell+1}, \ldots, t_{i-2})$ using {\sc RHBS}.
\end{algorithm}

As mentioned before, the value of $\alpha(r)$ is selected based on
the observation that (1) the probability that $A$ falls in the left
part of $T$ should be close to $1/2$ and (2) the length of the left
part for $r=1$ (which seems more important than other less
happening cases for $r \geq 2$) should be close to a power of two.  The previous selection is
perfect in terms of (1) but is not in terms of (2) since $\alpha(r)$ does not
depend on the length $i$ of $T$.  In this section, we put a priority
to (2) by setting
\begin{equation}\label{eq:alpha*}
\alpha(r, p_i)
=\left\{
\begin{array}{ll}
1 - \frac{1}{2^{k-1}} + \frac{1}{p_i2^{k+1}} \:\:& (r = 2k-1 \text{ and } p_i \in (3/4,1]),\\
1 - \frac{1}{2^k} - \frac{1}{p_i2^{k+2}} & (r = 2k-1 \text{ and } p_i \in (1/2,3/4]),\\
1 - \frac{1}{2^{k}} & (r = 2k).
\end{array}
\right.
\end{equation}

Note that it now depends on $i$ and it turns out that if $A$ falls
into the left part of $T$ for $r=1$, then the length of the left part
is exactly a power of two for any $i$ when $p_i \in (3/4,1]$. 
Note that for even $r$, $\alpha(r, p_i)$ is the same as the previous $\alpha(r)$.

We denote the modified {\sc 2Merge} as {\sc 2Merge$^*$} and
the whole sorting algorithm as {\sc (1,2)Insertion$^*$}. (See Algorithm)
Our analysis, having two cases for $ p_i \geq 3/4$
and $p_i < 3/4$, is more involved but we can obtain the average number
of comparisons for a single step is
\begin{eqnarray*}
\ceil{\lg i} + {\cal B}(i) \pm O(1/i)+ 
\left\{
\begin{array}{ll}
\frac{1}{2}-\frac{3}{4p_i}+\frac{25}{96p_i^2} \:\: & p_i \in (1/2, 3/4),\\
1-\frac{3}{2p_i}+\frac{13}{24p_i^2}& p_i \in [3/4,1].
\end{array}
\right.
\end{eqnarray*}

As with the previous section,
comparing this value with (\ref{eq:ins1}),
{\sc 2Merge}$^*$ is better than the binary insertion for 
$p_i \in \left[ \frac{3}{4} - \frac{\sqrt{6}}{12}, \frac{3}{4} + \frac{\sqrt{3}}{12}\right]$.
(Note that we did not use numerical analysis this time.)
Then, one step complexity of {\sc (1,2)Insertion}$^*$ is
\[
\ceil{\lg i} + {\cal D}^*(p_i)
\]
where
\[
{\cal D}^*(p_i) = \left\{
\begin{array}{ll}
1 - \frac{1}{p_i} & (1/2, \frac{3}{4} - \frac{\sqrt{6}}{12}],\\
\frac{3}{2}-\frac{7}{4p_i}+\frac{25}{96p_i^2} \:\:& p_i \in (\frac{3}{4} - \frac{\sqrt{6}}{12}, 3/4],\\
2 -\frac{5}{2p_i}+\frac{13}{24p_i^2}& p_i \in (3/4,\frac{3}{4} + \frac{\sqrt{3}}{12}],\\
1 - \frac{1}{p_i} & p_i \in (\frac{3}{4} + \frac{\sqrt{3}}{12},1].
\end{array}
\right.
\]

Thus, we have
\begin{eqnarray*}
\sum_{i=1}^{n}{\cal D}^*(p_i)=
2^{\ceil{\lg n}} \times \left\{
\int_{1/2}^{1}{\cal D}^*(x)dx
 + \int_{1/2}^{p_n}{\cal D}^*(x)dx
 \right\} + O(\log n)
\end{eqnarray*}
and can obtain the following theorem. 
(See Appendix \ref{app:imp2ins} for details.)

\begin{theorem}\label{thm:ins2*}
The complexity of {\sc (1,2)Insertion$^*$} is at most $n\lg n -1.4034n$.
\end{theorem}

We conducted an experiment for {\sc 2Merge$^*$}.  See Appendix~\ref{app:experiment}.
We prepare sequences $N=(1,2, \ldots, n)$ for $n$ up to $2^{12} =2046$.  Then
two elements $I_1$ and $I_2$ are selected from $N$ and they are
inserted into $N-\{I_1, I_2\}$ using {\sc 2Merge$^*$}.  
We take the average for the number of comparisons
for all possible pairs of $I_1$ and $I_2$.  As one can see the result
matches the analysis very well.  We also did a similar experiment for 
{\sc 2Merge}.  The result is very close and the difference is not
visible in such a graph.

%%%%%%%%%%%%%%%%%%%%%%%%%%%%%%%%%%%%%%%%%%%%%%%%%%%%%%%%%%%%%%
%% Section : Combined Algorithm
%%%%%%%%%%%%%%%%%%%%%%%%%%%%%%%%%%%%%%%%%%%%%%%%%%%%%%%%%%%%%%
\section{Combination with MergeInsertion}\label{sec:comb}

\begin{algorithm}[b]
\caption{ {\sc Combination}($S$)}
{\bf Input:} A (unsorted) sequence $S=(s_1, s_2, ..., s_n)$, where $n$ is even.\\
{\bf Output:} Sorted sequence\\
Step~1: If $p_n \geq 2/3$, then let $n':= \frac{2n}{3p_n}$. Otherwise, let $n':= \frac{n}{3p_n}$.\\
Step~2: Sort $S'=(s_1, ..., s_{n'})$ by the MergeInsertion sort to obtain $T'$.\\
Step~3: For $i = n'+2, n'+4, \ldots, n$,
if $p_i \in \left[ \frac{3}{4} - \frac{\sqrt{6}}{12}, \frac{3}{4} + \frac{\sqrt{3}}{12}\right]$,
then insert $s_{i-1}$ and $s_{i}$ into $T'$ by calling {\sc 2Merge*}$(s_{i-1}, s_{i}, T')$, 
otherwise insert $s_{i-1}$ into $T'$ by {\sc RHBS} and then $s_{i}$ by {\sc RHBS}.
\end{algorithm}

See Fig. 1, which illustrates the performance of 
{\sc (1,2)Insertion},
{\sc (1,2)Insertion}$^*$, and 
MergeInsertion~\cite{EW14} for the value of $p_n$.
As one can see, MergeInsertion 
is way better than our algorithms in a certain range of $p_n$.
In fact, due to \cite[page 389]{FJ59}, its best case happens for $n =
\ceil{\frac{2^k}{3}}$ for an integer $k$, achieving a complexity of
$n \lg n - (3-\lg 3) n + O(\lg n) \approx n \lg n - 1.415n + O(\lg n)$.
This best case can be easily included into our 
{\sc (1,2)Insertion$^*$}, as follows (see Algorithm~4):

Suppose that our input satisfies $p_n \geq 2/3$.  Then
we select the largest $k$ such that $n':= \ceil{\frac{2^k}{3}} \leq
n$.  Then we sort the first $n'$ elements by MergeInsertion.
After that the remaining elements are inserted by {\sc
  (1,2)Insertion$^*$}.  Since $n' = \frac{2n}{3p_n}$ as mentioned
  above, the complexity of MergeInsertion for that size is
\[
n' \lg n' - (3-\lg 3)n' = n' \ceil{\lg n} - \frac{4}{3p_n}n
\]
and the additional comparisons in {\sc (1,2)Insertion$^*$} cost is
\begin{eqnarray*}
\sum_{i = n'+1}^{n} \left\{ \ceil{\lg i} + {\cal D}^*(i) \right\}
= (n-n')\ceil{\lg n} + 2^{\ceil{\lg n}}\int_{2/3}^{p_n}{\cal D}^*(x)dx.
\end{eqnarray*}
Summing up these two quantities, we have
\begin{eqnarray*}
&&n'\ceil{\lg n} - \frac{4}{3p_n}n+(n-n')\ceil{\lg n} + 2^{\ceil{\lg n}}\int_{2/3}^{p_n}{\cal D}^*(x)dx\\
&=& 
n \ceil{\lg n} - \frac{4}{3p_n}n + 2^{\ceil{\lg n}}\int_{2/3}^{p_n}{\cal D}^*(x)dx\\
&=& 
n \lg n + \left\{-  \lg p_n - \frac{4}{3p_n} + \frac{1}{p_n}\int_{2/3}^{p_n}{\cal D}^*(x)dx \right\} n.
\end{eqnarray*}
We can use exactly the same approach for the case that $p_n \leq
2/3$.  It turns out however that the combined approach is worse than
MergeInsertion itself for $ 0.638 \leq p_n \leq 2/3$. So it is
better to use only MergeInsertion for this range.  See Fig. 1 for the
overall performance of the combined algorithm.

\begin{theorem}
The complexity of the combined algorithm is $n\lg n - 1.41064n$. 
\end{theorem}

\begin{figure}
\centering
\includegraphics[width=0.9\linewidth,clip]{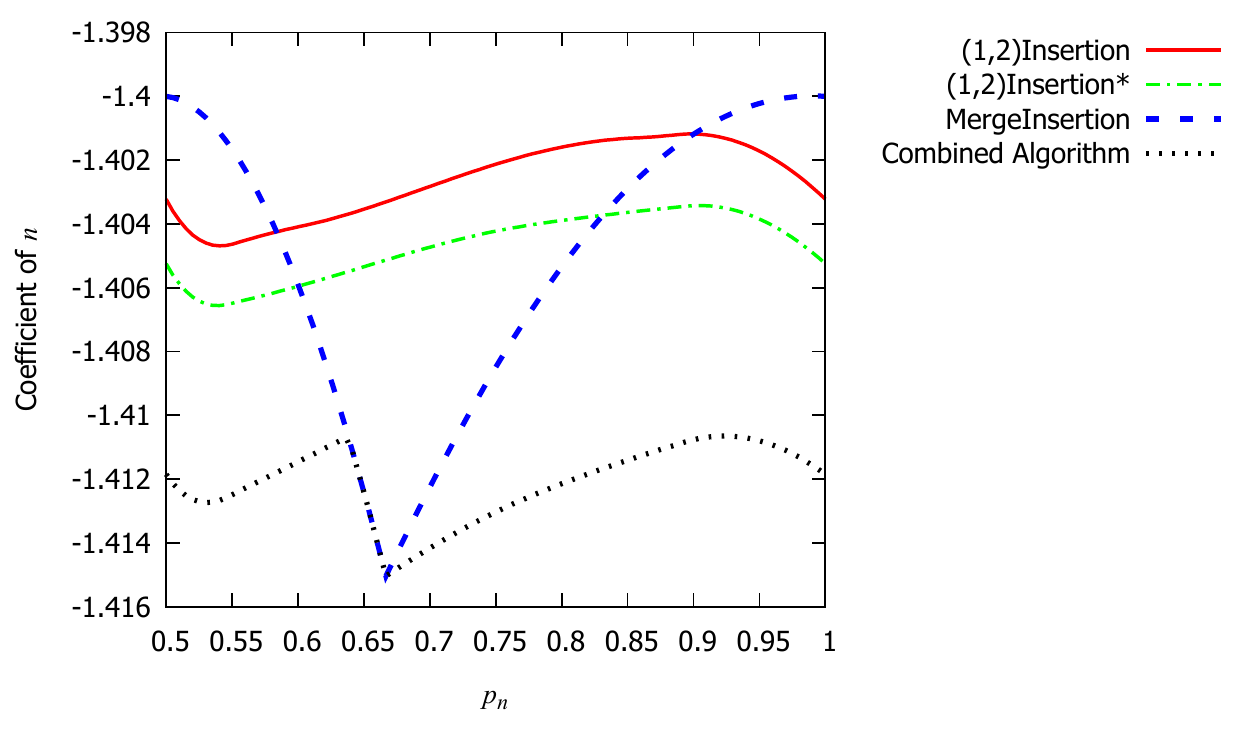}
\caption{Performance of the algorithms}
\end{figure}

\section{Final Remarks}

There is the wide agreement in the community that the
information-theoretic lower bound ($=-1.4427$) cannot be achieved by a
specific sorting algorithm; to prove or disprove it is a big open
question.  Anyway, our upper bound for the average case seems quite
close to the lower bound. So attacking the worst case using the ideas
in this paper may be more promising.

%%%%%%%%%%%%%%%%%%%%%%%%%%%%%%%%%%%%%%%
% Appendix
%%%%%%%%%%%%%%%%%%%%%%%%%%%%%%%%%%%%%%%
\newpage
\appendix
\section*{Appendix}

%%%%%%%%%%%%%%%%%%%%%%%%%%%%%%%%%%%%%%%
% Appendix : Analysis of E[r]
%%%%%%%%%%%%%%%%%%%%%%%%%%%%%%%%%%%%%%%
\section{Expected Number of Comparisons in Step~2}\label{app:expe_r}
Let $F_r$ be the event that $t_{\ceil{\alpha(r-1)\cdot i}}<A<t_{\ceil{\alpha(r)\cdot i}}$ holds similar to $F$ in Section~{2.1}.
Recall that $w_r = 2^{-r/2}(\sqrt{2}-1)i$.
Then, we have
$\prob[F_r] = \frac{w z}{2\binom{i}{2}}$,
where $w = w_r\pm1, z = 2^{-r/2}(\sqrt{2}+1)i\pm1$.
Then it turns out that $\prob[F_r]$ is close to $2^{-r/2}$ with an error term $O\left( \frac{1}{2^{r/2}i} \right)$,
in fact, \begin{eqnarray*}
\left| \frac{wz}{2\binom{i}{2}} - 2^{-r} \right| 
&=& 
\left| \frac{2^{-r}i^2 \pm 2(\sqrt{2}+1) \cdot  2^{-r/2} i \pm 1}{2\binom{i}{2}} - 2^{-r} \right| \\
&=& 
\left| \frac{1}{2^r (i-1)}\right| 
+ 
\left| \frac{\sqrt{2}+1}{(i-1)2^{\frac{r}{2}}} \right|
+ 
\left|\frac{1}{i(i-1)}
\right|\\
&=&O\left( \frac{1}{2^{r/2}i} \right) \quad ( \because 1 \leq r \leq 2\lg i)
\end{eqnarray*}
holds.
This implies that the expected value of $r$,
i.e., the number of comparisons at Step~2 
is 
\[
\expe[r] = \sum_{r=1}^{2\lg i} r \cdot Pr[F_r] = 2 \pm O\left(1/i\right).
\]

%%%%%%%%%%%%%%%%%%%%%%%%%%%%%%%%%%%%%%%
% Appendix : Analysis for Theorem 1
%%%%%%%%%%%%%%%%%%%%%%%%%%%%%%%%%%%%%%%
\section{Detailed Analysis for Theorem~1}\label{app:sum_int}

Theorem~1 is due to the following facts:
\begin{proposition}\label{fact:subsum_int}
For any integers $n$ and $n'$ such that $2^{m-1} \leq n < n' \leq 2^m$ hold and 
the $C^1$ function $f(x) : [1/2, 1] \rightarrow R$,
\begin{equation}\label{eq:subsum_int}
\sum_{i=n+1}^{n'} f(p_i) = 
2^{m} \cdot \int_{p_n}^{p_{n'}}f(x)dx \pm |f(p_{n'})-f(p_{n})|
\end{equation}
\end{proposition}
\proof
We use the trapezoidal rule:
For real values $a$ and $b$, an integer $N$, we have
\begin{eqnarray*}
\int_{a}^{b}f(x)dx = h \cdot \left\{ \frac{f(b)+f(a)}{2} + \sum_{i=1}^{N-1}f\left(a+k\cdot h\right)\right\}
- \frac{1}{12h^2}\cdot (f'(b) - f'(a)) + O(N^{-3})
\end{eqnarray*}
where $h=\frac{b-a}{N}$.
Setting $a=p_n$, $b=p_{n'}$ and $h=2^{-m}$ ($N = (p_{n'} - p_n)2^m = n'- n$), we have
\begin{eqnarray*}
\int_{p_{n}}^{p_{n'}}f(x)dx &=& \frac{1}{2^m} \left\{ \frac{f(p_{n})+f(p_{n'})}{2} + \sum_{i=1}^{n'-n-1}f\left( \frac{1}{2}+\frac{i}{2^m}\right)\right\}
+ O(1/n) \\
&=&\frac{1}{2^m} \left\{ \frac{f(p_n)+f(p_{n'})}{2} + \sum_{n+1}^{n'-1}f\left(\frac{i}{2^m}\right)\right\}
+ O(1/n) \\
&=&
\frac{1}{2^m} \sum_{i=n+1}^{n'}f\left(\frac{i}{2^m}\right)
+ \frac{f(p_n)-f(p_{n'})}{2^{m+1}} + O(1/n) \\
&=&
\frac{1}{2^m} \sum_{i=n+1}^{n'}f\left(p_i\right) + O(1/n),
\end{eqnarray*}
which means {(\ref{eq:subsum_int})}.
\qed

\begin{proposition}\label{fact:sum_int}
For any integer $n$ and the function $f(x) : [1/2, 1] \rightarrow R$,
\[
\sum_{i=1}^{n} f(p_i) = 
2^{\ceil{\lg n}} \cdot \left\{\int_{1/2}^{1}f(x)dx +  \int_{1/2}^{p_n}f(x)dx\right\} + O(\log n)
\]
holds.
\end{proposition}
\proof
Applying Proposition~\ref{fact:subsum_int},
\begin{eqnarray*}
\sum_{i=1}^{n} f(p_i) 
&=& 
\sum_{d=1}^{\ceil{\lg n}-1} \sum_{i=2^{d-1}+1}^{2^d} f(p_i) + \sum_{i=2^{\ceil{\lg n}-1}+1}^{n} f(p_i)\\
&=& 
\sum_{d=1}^{\ceil{\lg n}-1}\left\{ 2^{d} \cdot \int_{1/2}^{1}f(x)dx \pm \frac{|f(1)-f(1/2)|}{2^d} \right\}\\
&&\quad\quad
+ 2^{\ceil{\lg n}} \cdot \int_{1/2}^{p_n}f(x)dx \pm \frac{|f(p_n)-f(1/2)|}{2^{\ceil{\lg n}}}\\
&=&
2^{\ceil{\lg n}} \cdot \left\{\int_{1/2}^{1}f(x)dx +  \int_{1/2}^{p_n}f(x)dx\right\} + O(\log n)
\end{eqnarray*}
holds.
\qed

%%%%%%%%%%%%%%%%%%%%%%%%%%%%%%%%%%%%%%%
% Appendix : E[[r/2]]
%%%%%%%%%%%%%%%%%%%%%%%%%%%%%%%%%%%%%%%
\section{Expected values of $\ceil{r/2}$ and $\floor{r/2}$}\label{app:exp_ceil}
For an integer $r$, 
recall that $F_r$ is the event that Step~2 requires $r$ comparisons.
Also note that $\prob[F_r] = 2^{-r} \pm O\left( \frac{1}{2^{r/2}i} \right)$.

We ignore the $\pm O\left( \frac{1}{2^{r/2}i} \right)$ term for a while.
Namely, suppose that $\prob[F_r]$ is exactly $2^{-r}$.
Then, 
$\expe[\lfloor r/2 \rfloor] = \sum_{r=1}^{2\lg i} \floor{r/2}2^{-r}$.
The sum of terms with even $r$ can be written as 
\[
\sum_{t=1}^{\lg i} t 2^{-2t} = \frac{4}{9} - \frac{4}{9i^2} - \frac{\lg i}{3i^2} = \frac{4}{9} - O\left( \frac{\lg i}{i^2}\right)
\]
as using
\[
\sum_{k=1}^{\ell}kr^{k} = \frac{r}{(1-r)^2} - \frac{r^{\ell+1}}{(1-r)^2}- \frac{r^{\ell-1}\ell}{1-r}.
\]
For each odd $r$ term, $\floor{r/2} \cdot 2^{-r} = \frac{r-1}{2}2^{-(r-1)}\cdot 2^{-1}$.
This means that the sum of odd terms is a half of the sum of even ones.
Thus, if $\prob[F_r] = 2^{-r}$, then
\[
\expe[\lfloor r/2 \rfloor] = \frac{3}{2} \cdot \left(\frac{4}{9} - O\left( \frac{\lg i}{i^2}\right)\right)
= \frac{2}{3} - O\left( \frac{\lg i}{i^2}\right)
\]
holds. 
The error term $O\left(\frac{1}{2^{r/2}i}\right)$ of $\prob[F_r]$ is as much as $O(1/i)$ 
because
\[
\sum_{r=1}^{2\lg i} \floor{r/2} \cdot \frac{1}{2^{r/2}} < 
\sum_{r=1}^{\infty} \frac{r}{2^{r/2}} = O\left(1\right).
\]
Therefore, the expected value of $\floor{r/2}$ is close to 2/3 with an error
$O\left( \frac{1}{i}\right)$.

The rest of the proof is for the analysis of $\expe[\ceil{r/2}]$.
If $r$ is odd, then $\ceil{r/2} = \floor{r/2} + 1$, 
and $\ceil{r/2} = \floor{r/2}$ otherwise.
Adding the sum of $\prob[F_r]$ for odd $r$, we have
\[
\sum_{t = 1}^{\lg i} \left(\frac{1}{4^t} \pm O\left( \frac{1}{2^{t}i}\right)\right) = \frac{2}{3} - O\left( \frac{1}{i}\right),
\]
which means the expected value of $\floor{r/2}$ is close to 4/3 
with an error $O\left( \frac{1}{i}\right)$.
\qed

%%%%%%%%%%%%%%%%%%%%%%%%%%%%%%%%%%%%%%%
% Appendix : E[1/p] and E[1/p^2]
%%%%%%%%%%%%%%%%%%%%%%%%%%%%%%%%%%%%%%%
\section{Expected Values of $1/p_r$, $1/p_r^2$}\label{app:exp_pr}

Recall
\[
p_r 
= \frac{2^{-r/2}(\sqrt{2}-1)i}{2^{\lceil \lg i \rceil - c_r(p_i)}},
\text{ and }
c_r(p_i) :=
\left\{
\begin{array}{ll}
\lfloor r/2\rfloor + 2 \:\:&p_i \in (1/2, \frac{1+\sqrt{2}}{4}], \\
\lceil r/2\rceil + 1 \:\:&p_i \in (\frac{1+\sqrt{2}}{4}, \frac{2+\sqrt{2}}{4}], \\
\lfloor r/2\rfloor + 1 \:\:&p_i \in (\frac{2+\sqrt{2}}{4}, 1].
\end{array}
\right.
\]

Then, we have
\[
\frac{1}{p_r} =
\frac{\sqrt{2}+1}{p_i} \cdot \frac{2^{r/2}}{2^{c_r(p_i)}}, \:\:
\frac{1}{p^2_r}
=
\frac{3 +2\sqrt{2}}{p^2_i} \cdot \frac{2^{r}}{4^{c_r(p_i)}}.
\]

First, we give the expected value of $1/p_r$.
As with the proof of Lemma~\ref{lem:expect_ceilfloor}, 
let us suppose that $\prob[F_r]$ is exactly $2^{-r}$.
Our goal is to obtain
$\expe[ 2^{-\floor{r/2} + r/2}] = \sum_{r=1}^{2\lg i} 2^{-\floor{r/2} - r/2}$.
Considering the sum of even terms, we have
\begin{eqnarray*}
\sum_{t=1}^{\lg i} 2^{-2t} = \frac{1}{3} - O\left(\frac{1}{i}\right).
\end{eqnarray*}
For odd $r$, $2^{-\floor{r/2} - r/2} = 2^{-(r+1) + 3/2}$ holds.
This means that the sum of odd terms is $2\sqrt{2}$ times as the sum of even ones.
Then, we have
\begin{eqnarray*}
\expe[ 2^{-\floor{r/2} + r/2}] = \sum_{r=1}^{2\lg i} 2^{-\floor{r/2} -r/2} = \frac{1 + 2\sqrt{2}}{3} - O\left(\frac{1}{i}\right).
\end{eqnarray*}
Moreover, the error term is
$\sum_{r=1}^{2\lg i} 2^{-\floor{r/2} -r} \cdot O\left( \frac{1}{i}\right) = O\left(\frac{1}{i}\right).$
Therefore, when $p_i \in (1/2, (1+\sqrt{2})/4]$ 
\[
\expe[1/p_r] 
= \frac{\sqrt{2}+1}{p_i} \cdot \frac{\expe[ 2^{-\floor{r/2} + r/2}]}{4}
= \frac{5+3\sqrt{2}}{12p_1} - O\left(\frac{1}{i}\right),
\] and when $p_i \in ((1+\sqrt{2})/4, (2+\sqrt{2})/4]$ 
\[
\expe[1/p_r] 
= \frac{\sqrt{2}+1}{p_i} \cdot \frac{\expe[ 2^{-\floor{r/2} + r/2}]}{2}
= \frac{5+3\sqrt{2}}{6p_1} - O\left(\frac{1}{i}\right).
\]

Considering $\expe[ 2^{-\ceil{r/2} + r/2}] = \sum_{r=1}^{2\lg i} 2^{-\ceil{r/2} - r/2}$,
the difference from the `floor' case is that the sum of odd terms is $\sqrt{2}$ times the sum of even ones.
Then, we have
\begin{eqnarray*}\label{eq:ceil_term}
\expe[ 2^{-\ceil{r/2} + r/2}] = \frac{1 + \sqrt{2}}{3} - O\left(\frac{1}{i}\right).
\end{eqnarray*}
and 
when $p_i \in ((1+\sqrt{2})/4, (2+\sqrt{2})/4]$, 
\[
\expe[1/p_r] 
= \frac{\sqrt{2}+1}{p_i} \cdot \frac{\expe[ 2^{-\ceil{r/2} + r/2}]}{2}
= \frac{3+2\sqrt{2}}{6p_i} - O\left(\frac{1}{i}\right).
\]

Next, we give the expected value of $1/p^2_r$.
Assuming that $\prob[F_r]=2^{-r}$, 
we have $\expe[ 2^{-2\floor{r/2} + r}] = \sum_{r=1}^{2\lg i} 2^{-2\floor{r/2}}$.
The sum of even terms is $\frac{1}{3} - O(1/i)$.
The sum of odd terms is four times the sum of even ones.
Noting that the error term is also $O(1/i)$, we have
\[
\expe[ 2^{-2\floor{r/2} + r}] = \frac{5}{3} \pm O\left(\frac{1}{i}\right).
\]
It turns follows that when $p \in (1/2, (1 + \sqrt{2})/4]$
\[
\expe\left[1/p_r^2\right] 
= \frac{3+2\sqrt{2}}{p^2_i} \cdot \frac{\expe[ 2^{-2\floor{r/2} + r}]}{16}
= \frac{5(3 + 2\sqrt{2})}{48p^2_i} \pm O\left(\frac{1}{i}\right),
\]
and 
when $p_i \in ((2 + \sqrt{2})/4, 1]$
\[
\expe[1/p_r^2] 
= \frac{3+2\sqrt{2}}{p^2_i} \cdot \frac{\expe[ 2^{-2\floor{r/2} + r}]}{4}
= \frac{5(3 + 2\sqrt{2})}{12p^2_i}\pm O\left(\frac{1}{i}\right).
\]

Considering $\expe[ 2^{-2\ceil{r/2} + r}] = \sum_{r=1}^{2\lg i} 2^{-\ceil{r/2}}$,
the difference from `floor' case is that the sum of odd terms is the same as the sum of even ones.
Then, we have
\begin{eqnarray*}
\expe[ 2^{-2\ceil{r/2} + r}] = \frac{2}{3} \pm O\left(\frac{1}{i}\right).
\end{eqnarray*}
and 
when $p_i \in ((1+\sqrt{2})/4, (2+\sqrt{2})/4]$, 
\[
\expe[1/p_r^2] 
= \frac{3 + 2\sqrt{2}}{p^2_i} \cdot \frac{\expe[ 2^{-2\ceil{r/2} + r}]}{2}
= \frac{3+2\sqrt{2}}{6p^2_i} - O\left(\frac{1}{i}\right),
\]
which completes the proof.\qed

%%%%%%%%%%%%%%%%%%%%%%%%%%%%%%%%%%%%%%%%%%%%%%%%%%%%%%%%%%%%
% Appendix : Analysis for Improved Algorithm
%%%%%%%%%%%%%%%%%%%%%%%%%%%%%%%%%%%%%%%%%%%%%%%%%%%%%%%%%%%%
\section{Proof of Theorem~2}\label{app:imp2ins}

In this section, we give the average number of 
comparisons that {\sc 2Merge}$^*$ requires at Step~2 and Step~3.
There are two cases:(i) $p_i \in (3/4, 1]$ and (ii) $p_i \in (1/2, 3/4]$.

\subsection{Case I: $p_i \in (3/4, 1]$}

Our goal is to show that the average number of comparisons 
for insertion of $A$ is
\[
\ceil{\lg i} + 2 - \frac{3}{p_i} + \frac{3}{4p_i^2} \pm O\left(\frac{1}{i}\right).
\]

If Step~2 has made $r$ comparisons,
we call {\sc RHBS}($A$, $t_{\ceil{\alpha(r-1)i}+1}, ..., t_{\ceil{\alpha(r)i}-1}$).
We denote by $w_r$ the number of candidate positions for insertion, 
that is, $w_r = \ceil{\alpha(r,p_i)i} - \ceil{\alpha(r-1,p_i)i}$.

For the case of $r=2k-1$, because 
\[
w_{2k-1} = \alpha(r,p_i)i -\alpha(r-1,p_i)i \pm 1  = 2^{\ceil{\lg i}-k-1} \pm 1
\]
and $w_{2k-1}$ is an integer, 
we have $w_{2k-1} = \frac{i}{p_i \cdot 2^{k+1}} = 2^{\ceil{\lg i}-k-1}$.
Since $w_{2k-1}$ is the power of two, 
{\sc RHBS} requires $\ceil{\lg i}-k-1$ comparisons.

For the case of $r=2k$, we have
\[
w_{2k}
=
\frac{i}{2^k} \cdot \left( 1 - \frac{1}{2p_i}\right) \pm 1 = 2^{\ceil{\lg i}-k-1} \cdot \left(2p_i - 1 \right) \pm 1.
\]
The value $\ceil{\lg w_{2k}}$ is obviously $\ceil{\lg i}-k-2$ or $\ceil{\lg i}-k-1$.
Let us denote $i = 3 \cdot 2^{\ceil{\lg i}-2} + \beta$, where $\beta \in [1, 2^{\ceil{\lg i}-2}]$.
Then, we also have 
\begin{eqnarray*}
w_{2k} &=& \ceil{ i-\frac{i}{2^k}} - \ceil{ i-\frac{i}{2^{k-1}} + \frac{i}{p_i2^{k+1}} }\\
&=& \ceil{-3 \cdot 2^{\ceil{\lg i}-k-2} - \frac{\beta}{2^k}} - \ceil{ -2^{\ceil{\lg i}-k} - \frac{2\beta}{2^k} } \\
&=& 2^{\ceil{\lg i}-k-2} + \floor{ \frac{2\beta}{2^k} }- \floor{\frac{\beta}{2^k}}.
\end{eqnarray*}
Therefore, $w_{2k} = 2^{\ceil{\lg i}-k-2}$, that is, $\lg w_{2k} = \ceil{\lg i}-k-2$ if $\beta < 2^{k-1}$,
and $\ceil{\lg w_{2k}} = \ceil{\lg i}-k-1$ otherwise.
When $\ceil{\lg w_{2k}} = \ceil{\lg i} - k - 1$, we have
\[
\frac{2^{\ceil{\lg w_{2k}}}}{w_{2k}} 
= \frac{2^{\ceil{\lg i} - k - 1}}{2^{\ceil{\lg i}-k-1} \cdot \left(2p_i - 1 \right) \pm 1}
= \frac{1}{ 2p_i - 1} \pm O\left( 2^k/i \right),
\]
and when $\ceil{\lg w_{2k}} = \ceil{\lg i} - k - 2$, we have
\[
\frac{2^{\ceil{\lg w_{2k}}}}{w_{2k}} 
= \frac{1}{ 4p_i - 2} \pm O\left( 2^k/i \right).
\]
Let $z_{r}=2i-\ceil{\alpha(r-1,p_i)\cdot i}-\ceil{\alpha(r,p_i)\cdot i}-1$.
Since $z_{2k} = 2^{\ceil{\lg i}-k-1} \cdot \left(6p_i - 1 \right) \pm 1$,
we have
\[
\frac{w_{2k}}{z_{2k}} 
= \frac{2p_i-1}{6p_i-1} \pm O\left( 2^k/i \right).
\]

As the proof of Lemma~\ref{lem:step3}, the average number of comparisons that {\sc RHBS} requires for the case of $r=2k$ is 
\[
\Delta(2k) = \ceil{\lg w_{2k}} - \left(\frac{2^{\ceil{\lg w_{2k}}}}{w_{2k}}-1 \right)
\cdot \left( 1 + \left( 2 - \frac{2^{\ceil{\lg w_{2k}}}}{w_{2k}} \right) \cdot \frac{w_{2k}}{z_{2k}} \right).
\]

When $\ceil{\lg w_{2k}} = \ceil{\lg i} - k - 1$, 
\begin{eqnarray*}
\Delta(2k)&=& 
\ceil{\lg w_{2k}} -
\frac{2-2p_i}{2p_i-1}
\cdot \left( 1 + \frac{4p_i-3}{2p_i-1} \cdot \frac{2p_i-1}{6p_i-1}\right)  \pm O(2^k/i)\\
&=&
\ceil{\lg w_{2k}} - \frac{2-2p_i}{2p_i-1}\cdot \frac{10p_i-4}{6p_i-1} \pm O(2^k/i)\\
&=& 
\ceil{\lg i} - k - 2 + 
\frac{(8p_i -3)(4p_i -3)}{(2p_i-1)(6p_i-1)} \pm O(2^k/i).
\end{eqnarray*}

When $\ceil{\lg w_{2k}} = \ceil{\lg i} - k - 2$, 
$i < 3 \cdot 2^{\ceil{\lg i}-2} + 2^{k-1}$ holds.
Then, $p_i = \frac{3}{4} + O(2^k/i)$.
Because 
\[
\frac{2^{\ceil{\lg w_{2k}}}}{w_{2k}}-1 = \frac{1}{4p_i-2}-1 \pm O\left(2^k/i\right) 
= \frac{4p_i-3}{4p_i-2} \pm O\left(2^k/i\right) = \pm O\left(2^k/i\right)
\]
holds, we have
$\Delta(2k) = \ceil{\lg i} - k - 2 \pm O(2^k/i)$.
Note that we can deal with both cases as
$\ceil{\lg i} - k - 2 + \frac{(8p_i -3)(4p_i -3)}{(2p_i-1)(6p_i-1)} \pm O(2^k/i).$

Let us denote by $F_r$ the event that $t_{\ceil{\alpha(r-1,p_i)\cdot i}}<A<t_{\ceil{\alpha(r,p_i)\cdot i}}$ holds.
Then, we have
\begin{eqnarray*}
\prob\left[F_{r}\right]
=
\sum_{\ell=\ceil{\alpha(r-1,p_i)\cdot i}+1}^{\ceil{\alpha(r, p_i)\cdot i}} \frac{i-\ell}{\binom{i}{2}}
=
\frac{w_{r} z_{r}}{2\binom{i}{2}}
=
\left\{
\begin{array}{lc}
\frac{8p_i-1}{p_i^2 4^{k+1}} \pm O\left(\frac{1}{i\cdot 2^{k}}\right) & r=2k-1,\\
\frac{(2p_i-1)(6p_i+1)}{4^{k+1}p_i^2} \pm O\left(\frac{1}{i\cdot 2^k}\right) & r=2k.
\end{array}
\right.
\end{eqnarray*}
Therefore, we have
\begin{eqnarray*}
\sum_{k=1}^{\lg i} \prob\left[F_{2k}\right] = \frac{(2p_i-1)(6p_i-1)}{12p_i^2}\pm O\left(\frac{1}{i}\right),
\end{eqnarray*}
and 
\begin{eqnarray*}
\prob\left[F_{2k-1}\right] + \prob\left[F_{2k}\right] = \frac{3}{4^k}\pm O\left(\frac{1}{i2^k}\right).
\end{eqnarray*}

Thus, the average number of comparisons that Step~2 and Step~3 require is\\

\begin{eqnarray*}
&&\sum_{k \geq 1}^{} \prob[F_{2k-1}] \times \left\{ 2k -1 + \ceil{\lg i} - k - 1\right\}
\\&&
\quad\quad+\sum_{k \geq 1}^{} \prob[F_{2k}] \times \left\{ 2k + \ceil{\lg i} - k - 2 + \frac{(8p_i -3)(4p_i -3)}{(2p_i-1)(6p_i-1)} \pm O(2^k/i) \right\}\\
&=& \ceil{\lg i} - 2
+ \sum_{r \geq 1}^{} k \cdot \left\{ \prob[F_{2k-1}] + \prob[F_{2k}] \right\}
+\sum_{k \geq 1}^{} \prob[F_{2k}] \times \left\{ \frac{(8p_i -3)(4p_i -3)}{(2p_i-1)(6p_i-1)} \pm O(2^k/i) \right\} \\
&=& \ceil{\lg i} - 2
+ \sum_{k \geq 1}^{} \frac{3k}{4^k} + \frac{(2p_i-1)(6p_i-1)}{12p_i^2} \times \frac{(8p_i -3)(4p_i -3)}{(2p_i-1)(6p_i-1)} + O\left(\frac{1}{i}\right)\\
&=& \ceil{\lg i} + 2 - \frac{3}{p_i} + \frac{3}{4p_i^2} \pm O\left(\frac{1}{i}\right).
\end{eqnarray*}

\subsection{Case II: $p_i \in (1/2, 3/4]$}

We show that the average number of comparisons Step~3 requires is
\[
\ceil{\lg i} + 1 - \frac{3}{2p_i} + \frac{3}{16p_i^2} \pm O\left(\frac{1}{i}\right).
\]

Since $p_i < 3/4$ and $i \geq 6$ is even, we have $\ceil{ \lg i } \geq 3$.

First, we give the average number of comparisons at Step~3 when $r=1$.
In this case, we call {\sc RHBS}($A$, $(t_1, \ldots, t_{\ceil{\alpha(1, p_i)i}-1})$).
Let $w = \ceil{\alpha(1, p_i)i}$.
Then, we have
\begin{eqnarray*}
w = \ceil{\frac{i}{2} - \frac{i}{8p_i}} 
=2^{\ceil{\lg i}}\cdot\frac{4p_i - 1}{8}.
\end{eqnarray*}
This implies that $2^{\ceil{\lg i}-3} < w \leq 2^{\ceil{\lg i}-2}$, 
that is, $\ceil{\lg w} = \ceil{\lg i} -2$ holds since $p_i \in (1/2, 3/4]$.
Let $z = 2i - w - 1$.
The average number of comparisons that {\sc RHBS} requires is 
\[
\Delta = \ceil{\lg w} - 
\left( \frac{2^{\ceil{\lg w}}- w}{w}\right) \cdot
\left( 1 + \frac{w-2^{\ceil{\lg w}}}{w} \cdot 
\frac{w}{z}\right),
\]
because the following holds:
\[
\frac{2^{\ceil{\lg w}}}{w} = \frac{2}{4p_i-1} \pm O(1/i), \:\:
\frac{w}{z} = \frac{4p_i-1}{12p_i+1} \pm O(1/i),
\]
\begin{eqnarray*}
\Delta &=&
\ceil{\lg w} - \frac{-4p_i+3}{4p_i-1} \cdot \left( 1 + \frac{8p_i-4}{4p_i-1} \cdot \frac{4p_i-1}{12p_i+1}\right) \pm O(1/i)\\
&=& 
\ceil{\lg w} + \frac{(4p_i-3)(20p_i-3)}{(4p_i-1)(12p_i+1)} \pm O(1/i)\\
&=& 
\ceil{\lg i} + \frac{-16p_i^2-56p_i+11}{(4p_i-1)(12p_i+1)} \pm O(1/i).
\end{eqnarray*}

We consider the case of $t_{w}<A$, that is, $r>1$.
Let $i' = i - w = 2^{\ceil{\lg i}}\frac{4p_i + 1}{8}$.
Because 
$\frac{3}{8} \cdot 2^{\ceil{\lg i}} < i' \leq 2^{\ceil{\lg i}-1}$, 
$\ceil{\lg i'} = \ceil{\lg i} -1$ holds.
Moreover, we have
\[
p_{i'} = \frac{i'}{2^{\ceil{\lg i'}}} = p_i + \frac{1}{4}
\in (3/4, 1].
\]

If $t_{w}<A$, the operations of Step~2 and Step~3 after the first comparison 
is equivalent to the case that our insertion goes to length $i'-2$ sequence.
Because $p_{i'} \in (3/4, 1]$, we can apply the result of Case~I.
Then, the average number of Step~2 and Step~3 after the first comparison is
\[
\ceil{\lg i'} - \frac{3}{p_{i'}} + \frac{3}{4p_{i'}^2} 
\pm O\left(\frac{1}{i'}\right) 
=
\ceil{\lg i} + 1 - \frac{12}{4p_i+1} + \frac{12}{(4p_i+1)^2} 
\pm O\left(\frac{1}{i}\right).
\]

As with Case~I, we have
\[
\prob[F_1] = \frac{(4p_i-1)(12p_i+1)}{64p_i^2} \pm O(1/i),
\]
and 
\[
\prob\left[\bigcup_{r>1} F_r\right] = 1- \prob[F_1] = \frac{(4p_i+1)^2}{64p_i^2} \pm O(1/i).
\]

Therefore, adding the cost for the first comparison, 
the average number of comparisons at Step~2 and Step~3 is
\begin{eqnarray*}
&&1 + \prob[F_1] \times \left\{\ceil{\lg i} + \frac{-16p_i^2-56p_i+11}{(4p_i-1)(12p_i+1)} \right\} +
\prob\left[\bigcup_{r>1} F_r\right] \times \left(\ceil{\lg i} + 1 - \frac{12}{4p_i+1} + \frac{12}{(p_i+1)^2}\right)\\
&=& \ceil{\lg i} + 1 +
\frac{-16p_i^2-56p_i+11}{64p_i^2} 
+ \frac{(4p_i+1)^2}{64p_i^2} - \frac{3p_i+3}{16p_i^2} + \frac{3}{16p_i^2} \pm O(1/i)\\
&=& \ceil{\lg i} + 1 - \frac{3}{2p_i} + \frac{3}{16p_i^2}  \pm O(1/i).
\end{eqnarray*}

\subsection{Average Number of Comparisons for {\sc 2Merge}$^*$}
From the above arguments, adding the average number of comparisons for Step~1 and Step~4, 
the new average number of comparisons for inserting $i$-th elements 
is 
\begin{eqnarray*}
\ceil{\lg i} + {\cal B}(i) \pm O(1/i)+ 
\left\{
\begin{array}{ll}
\frac{1}{2}-\frac{3}{4p_i}+\frac{25}{96p_i^2} \:\: & p_i \in (1/2, 3/4],\\
1-\frac{3}{2p_i}+\frac{13}{24p_i^2}& p_i \in (3/4,1].
\end{array}
\right.
\end{eqnarray*}

Comparing with $\ceil{\lg i} + {\cal B}(i)$, 
{\sc 2Merge}$^*$ is better than the standard binary insertion
when $p_i \in \left[ \frac{3}{4} - \frac{\sqrt{6}}{12}, \frac{3}{4} + \frac{\sqrt{3}}{12}\right]$.
Then, one step complexity of {\sc (1,2)Insertion}$^*$ is
\[
\ceil{\lg i} + {\cal D}^*(p_i)
\]
where
\[
{\cal D}^*(p_i) = \left\{
\begin{array}{ll}
1 - \frac{1}{p_i} & (1/2, \frac{3}{4} - \frac{\sqrt{6}}{12}],\\
\frac{3}{2}-\frac{7}{4p_i}+\frac{25}{96p_i^2} \:\:& p_i \in (\frac{3}{4} - \frac{\sqrt{6}}{12}, 3/4],\\
2 -\frac{5}{2p_i}+\frac{13}{24p_i^2}& p_i \in (3/4,\frac{3}{4} + \frac{\sqrt{3}}{12}],\\
1 - \frac{1}{p_i} & p_i \in (\frac{3}{4} + \frac{\sqrt{3}}{12},1].
\end{array}
\right.
\]

As Section~2, we have
\begin{eqnarray*}
\sum_{i=1}^{n}{\cal D}^*(p_i)=
2^{\ceil{\lg n}} \times \left\{
\int_{1/2}^{1}{\cal D}^*(x)dx
 + \int_{1/2}^{p_n}{\cal D}^*(x)dx
 \right\} + O(\log n)
\end{eqnarray*}
and we obtain Theorem~\ref{thm:ins2*}.

\section{Experiments for {\sc 2insertion}}\label{app:experiment}
See Fig.2, which illustrates our analysis and 
results of simulations.
The symbol `+' means 
the average number of comparisons
of simulation for each $n$.

The line represents the value of analysis:
\begin{eqnarray*}
\ceil{\lg i} + {\cal B}(i) + 
\left\{
\begin{array}{ll}
\frac{1}{2}-\frac{3}{4p_i}+\frac{25}{96p_i^2} \:\: & p_i \in (1/2, 3/4),\\
1-\frac{3}{2p_i}+\frac{13}{24p_i^2}& p_i \in [3/4,1].
\end{array}
\right.
\end{eqnarray*}

We prepare sequences $N=(1,2, \ldots, n)$ for $n$ up to $2^{12} =2046$.  Then
two elements $I_1$ and $I_2$ are selected from $N$ and they are
inserted into $N-\{I_1, I_2\}$ using {\sc 2Merge$^*$}.  
We take the average for the number of comparisons
for all possible pairs of $I_1$ and $I_2$.  As one can see the result
matches the analysis very well.  We also did a similar experiment for 
{\sc 2Merge}.  The result is very close and the difference is not
visible in such a graph.

\begin{figure}
\centering
%\begin{center}
\includegraphics[width=10cm,clip]{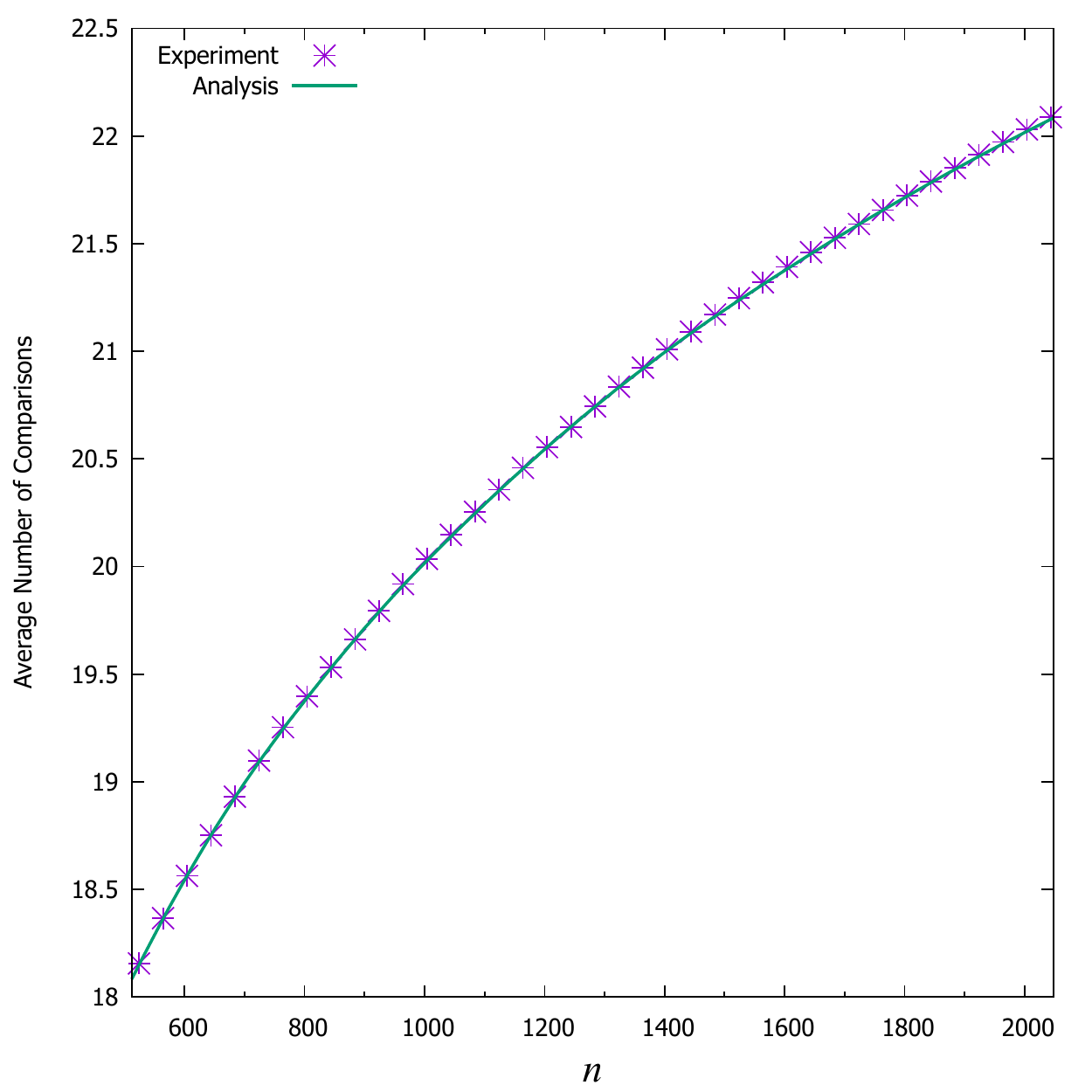}
\caption{
The average number of comparisons of {\sc 2Merge}: Experiment and analysis}
%\end{center}
\end{figure}


\begin{thebibliography}{99}

\bibitem{ayala2007}
Ayala-Rinc{\'o}n, M., De~Abreu, B.~T. and De~Siqueira, J.: A variant of the Ford--Johnson algorithm that is more space efficient, 
{\em Information Processing Letters\/},  Vol.~102, No.~5, pp.\ 201--207 (2007).

\bibitem{EW14}
S.~Edelkamp and A.~Wei\ss. QuickXsort: Efficient Sorting with n logn - 1.399n + o(n) Comparisons on Average. CSR 2014: pp. 139--152.

\bibitem{FJ59}
Ford, L.~R. and Johnson, S.~M.: A tournament problem, {\em The American
  Mathematical Monthly\/},  Vol.~66, No.~5, pp.\ 387--389 (1959).

\bibitem{HL71}
F.~K.~Hwang and S.~Lin, Optimal merging of 2 elements with n elements, 
Acta Informatica 1, pp.145--158, 1971.

\bibitem{Knuth98}
Knuth, D.~E.: {\em The Art of Computer Programming, Volume 3: (2nd Ed.) Sorting
  and Searching\/}, Addison Wesley Longman Publishing Co., Inc., Redwood City,
  CA, USA (1998).

\bibitem{mana79}
G.K. Manacher, The Ford--Johnson algorithm is not optimal,
{\em Journal of the Association for Computing Machinery\/} 26 (1979)
441--456.

\bibitem{mana89}
G.K. Manacher, T.D. Bui, T. Mai, Optimum combinations of
sorting and merging, {\em Journal of the Association for Computing
Machinery\/} 36 (1989) 290--334.

\bibitem{mon81}
J. Schulte M\"{o}nting, Merging of 4 or 5 elements with n elements,
{\em Theoretical Computer Science\/} 14 (1981) 19--37.

\bibitem{pec2002}
Peczarski, M.: Sorting 13 Elements Requires 34 Comparisons. In
Proc. 10th Annual European Symposium on Algorithms.
LNCS, vol. 2461, pp. 785--794. Springer (2002).

\bibitem{pec2004}
Peczarski, M.: New results in minimum-comparison sorting, {\em Algorithmica\/},
   Vol.~40, No.~2, pp. 133--145 (2004).

\bibitem{pec2007}
Peczarski, M.: The Ford--Johnson algorithm still unbeaten for less than 47
  elements, {\em Information processing letters\/},  Vol.~101, No.~3, pp.\
  126--128 (2007).

\bibitem{Steinhaus50}
H.~Steinhaus, Mathematical Snapshots, New Nork, 1950, pp. 37--40.

\bibitem{TAB86}
M.~Thanh, V.S.~Alagar, T.~D.~Bui, 
Optimal Expected-Time Algorithms for Merging. J. Algorithms 7(3): pp. 341--357, 1986.

\bibitem{well66}
Wells, M.: Applications of a Language for Computing in Combinatorics. 
In Proc. 1965 IFIP Congress,
North-Holland, Amsterdam, pp. 497--498 (1966).

\end{thebibliography}
\end{document}